\documentclass[aps,prl,reprint,showpacs,superscriptaddress]{revtex4-1} 
\usepackage{times}
\usepackage{graphicx}

\begin{document}

\title{Highly tunable low-threshold optical parametric oscillation \\
 in radially poled whispering gallery resonators}

\author{T. Beckmann}
\email{beckmann@uni-bonn.de}
\author{H. Linnenbank}
\author{H. Steigerwald}
\affiliation{Institute of Physics, University of Bonn, Wegelerstrasse 8, 53115 Bonn, Germany}
\author{B. Sturman}
\affiliation{Institute for Automation and Electrometry of RAS, 630090 Novosibirsk, Russia}
\author{D. Haertle}
\affiliation{Institute of Physics, University of Bonn, Wegelerstrasse 8, 53115 Bonn, Germany}
\author{K. Buse}
\affiliation{Institute of Microsystem Technology (IMTEK), University Freiburg, Georges-K\"ohler-Allee 102, 79110 Freiburg, Germany}
\affiliation{Fraunhofer Institute of Physical Measurement Techniques, Heidenhofstra{\ss}e 8, 79110 Freiburg, Germany}
\author{I. Breunig}
\affiliation{Institute of Physics, University of Bonn, Wegelerstrasse 8, 53115 Bonn, Germany}

\begin{abstract}
Whispering gallery resonators (WGR's), based on total internal
reflection, possess high quality factors in a broad spectral range. Thus,
nonlinear optical processes in such cavities are ideally suited for
the generation of broadband or tunable electromagnetic radiation.
Experimentally and theoretically, we investigate the tunability of
optical parametric oscillation in a radially structured WGR made of
lithium niobate. With a $1.04$-$\mu$m pump wave, the signal and
idler waves are tuned from $1.78$ to $2.5~\mu$m -- including the
point of degeneracy -- by varying the temperature between $20$ and
$62^{\,\circ}$C. A weak off-centering of the radial domain structure
extends considerably the tuning capabilities. The oscillation
threshold lies in the mW-power range.

\end{abstract}


\maketitle

Optical micro-resonators (micro-cavities) attract increasing
research interest owing to their outstanding properties and promises
for numerous applications~\cite{Vahala03, Del07, Carmon07,
Spillane03}. Ultra-high quality factors of optical modes, reaching
$Q \approx 10^{11}$~\cite{GrudinPR06}, together with small mode
volumes provide unprecedented conditions for shaping and enhancement
of light-matter interactions. Already existing and highly demanded
applications of these fascinating new optical elements span from
quantum electrodynamics and optics to optical filters and
sensors~\cite{Vahala03,Spillane03,Liang10}.

Bringing nonlinear-optical effects to the range of low-power
continuous-wave (cw) light sources is one of the biggest challenges.
Whispering gallery resonators (WGR's), made of crystalline nonlinear
media, are best suited for this
purpose~\cite{GrudinPR06,SavchenkovOE07,SavchenkovPR04}. The quality
factors, which are restricted from above by light absorption, are
among the highest ones. Together with a $\mu$m-size transversal
confinement, a $\sim10^7$ intensity enhancement can be reached. The
discrete WGR mode structure is well-understood nowadays and the
techniques for its engineering and coupling light in and out are
well developed~\cite{Matsko06,Gorodetsky06,StrekalovOL09}.

For most of the nonlinear-optical phenomena, phase matching is a key
issue. In WGR's it acquires specific features: While the most
excitable equatorial and near-equatorial modes can be treated as
plane waves, one has to take into account a discrete
geometry-dependent mode structure as well as a modified dispersion
relation~\cite{Matsko06}. The main impact of the discreteness is
here in the reduced possibility to meet phase matching. The smaller
the WGR size, the stronger is this impact~\cite{IlchenkoJOSA03}.
Usually, the fine adjustment to narrow-band cavity nonlinear
resonances can be made by varying the temperature or applying an
electrical field~\cite{FurstPRL10}.

A number of important nonlinear effects have been realized in WGR's
with low-power cw light sources during the last
years~\cite{Kieu07,Carmon05,Carmon07,SavchenkovPRL04,Kippen04,Del07,IlchenkoPRL04,FurstPRL10,Sasagawa09,Furst10}.
In $\chi^{(3)}$ optical materials, Raman scattering and lasing,
third-harmonic generation, low-threshold optical oscillation owing
to resonant four-wave mixing, and optical comb generation via Kerr
nonlinearity were reported. In $\chi^{(2)}$ materials,
second-harmonic generation, third harmonic generation by cascading
two second-order processes, and parametric down conversion were
demonstrated recently.

Here we tackle the fundamental problem of tunability of nonlinear
processes in WGR's by indicating how to vary optical frequencies
over wide ranges. Specifically, we demonstrate for the first time a
highly tunable low-threshold optical parametric oscillation (OPO) in
a radially poled WGR made of lithium niobate ($\text{LiNbO}_3$).

\begin{figure}[b]
\centering
\includegraphics[width=8.1cm]{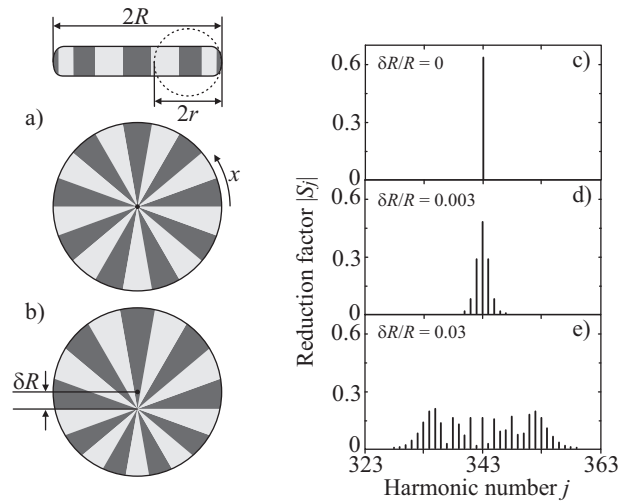}
\caption{Geometric properties of radially poled WGR's: Ideal and
distorted radial domain structures -- a) and b) -- and
representative spectra of the reduction factor $|S_j|$ -- c), d),
and e) -- for $2N = 686$ and $\delta R/R = 0$, $0.003$, and $0.03$,
respectively, as defined by Eq.~(1). Light and dark gray regions in
a) and b) indicate the alternating domains, $x$ is the circumference
coordinate, $R$ and $r$ are the large and small WGR
radii.}\label{Fig1}
\end{figure}

As it is known from the studies of $\chi^{(2)}$ nonlinear-optical
effects in bulk crystals, quasi-phase matching via domain
engineering~\cite{Fejer92} provides great scope for
tuning~\cite{Dunn99}. In the WGR case, a radial structure with an
even number $2N \gg 1$ of domains, see Fig.~1a, is best suited for
that purpose. For the desired nonlinear process, the optimal large
radius~$R$ at a specific domain number can be evaluated taking into
account the geometry-dependent corrections to refractive indices of
the interacting waves~\cite{Gorodetsky06}.

In polar optical materials, including lithium niobate, domain
inversion results in changing the sign of the nonlinear
susceptibility~$\chi^{(2)}$. To account for the $2\pi R$ periodicity
of the WGR case, we represent the sign-changing distribution
$\chi^{(2)}(x)$, where $x$ is the circumference coordinate, by the
Fourier series
\begin{equation}\label{Fourier}
S(x) = \sum\limits_{j = -\infty}^{\infty}
S_j\,\exp\Big(\mathrm{i}\,\frac{jx}{R} \Big)\,, \qquad   S_j = S_j^*
\,,
\end{equation}
\noindent where the Fourier coefficients are given by the average
over the circumference, $S_j = \langle
S(x)\,\exp(-\mathrm{i}jx/R)\rangle$, and $S(x) = {\rm
sign}[\chi^{(2)}(x)]$. The $1/R$ separation between the spatial
frequencies, dictated by the $2\pi R$ periodicity, is indeed the
same as the wavevector difference for the equatorial modes. As soon
as the domain pattern (the points of the sign change) is known, one
can calculate $S_j$ numerically. The quantity $|S_j|$ represents the
reduction factor for the nonlinear coefficient when using the
$j$-harmonic; it depends solely on the WGR geometry. If the domain
pattern is strictly periodic, the Fourier components $S_j$ can
easily be calculated analytically. Only the harmonics with $j = \pm
N, \pm 3N, \ldots$ are nonzero in this case, and the reduction
factor for the main harmonic, $|S_N| = 2/\pi$, is close to~$1$. Any
distortion of the periodicity of the domain structure makes nonzero
all spatial harmonics~$S_j$.

In our case, the most significant distortion originates from
off-centering of the radial structure, see Fig.~1b. It gives close
side harmonics with $j = N \pm 1,\, N \pm 2$, etc. Subfigures~1c) to
1e) illustrate the impact of off-centering on the Fourier spectrum
of the reduction factor $|S_j|$ for $2N = 686$, which is relevant to
our experiment. One sees that the first side harmonics become
comparable with the main one already for $\delta R/R = 0.003$. For
$\delta R/R = 0.03$, we have a rich discrete spectrum, where the
main harmonic is not the dominating one. The reduction factor
remains pretty large for many spatial harmonics. Failure in the
inversion of part of the domains is not critical for the spectrum,
because the long-range order is maintained.

Each $S_j$-peak can generally be used for phase matching. In the OPO case, the
corresponding phase-matching conditions read
\begin{equation}\label{PMC1}
\frac{1}{\lambda_\mathrm{p}} = \frac{1}{\lambda_\mathrm{s}} +
\frac{1}{\lambda_\mathrm{i}} \,, \qquad
\frac{n_\mathrm{p}}{\lambda_\mathrm{p}} =
\frac{n_\mathrm{s}}{\lambda_\mathrm{s}} +
\frac{n_\mathrm{i}}{\lambda_\mathrm{i}} + \frac{j}{2\pi R} \,,
\end{equation}
\noindent where $\lambda_\mathrm{p,s,i}$ are the pump, signal, and idler wavelengths.
The effective refractive index $n$, entering these relations, is a known function of
$\lambda$, $T$, $R$, and $r$~\cite{Edwards84,Gorodetsky06}.

\begin{figure}[hb]
\centering
\includegraphics[width=8.1cm]{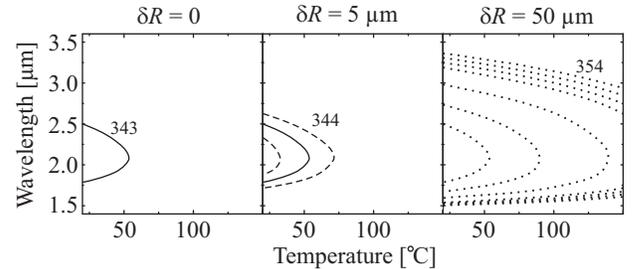}
\caption{Tuning curves $\lambda_\mathrm{s,i}(T)$ for $\lambda_\mathrm{p} = 1.04~\mu$m,
$2N = 686$, $R = 1.54$~mm, $r = 0.6$~mm, and $\delta R/R = 0$, $0.003$, and $0.03$. The
solid lines correspond to the strongest spectral peaks with $|S_j| > 0.3$, the dashed
lines to $0.2 < |S_j| \leq 0.3$, and the dotted lines to $0.1 < |S_j| \leq 0.2$. The
curves corresponding to $|S_j| < 0.1$ are dropped. }\label{Fig2}
\end{figure}

Figure~2 shows tuning curves calculated numerically from Eqs.~(\ref{PMC1}) for a pump
wavelength $\lambda_\mathrm{p} = 1.04~\mu$m, the WGR radii $R = 1.54$~mm and $r =
0.6$~mm, extraordinary polarization for all three waves, and several values of $j$.
Remarkably, there is a substantial difference in the tuning curves even for neighboring
$S_j$-peaks. In essence, the tuning possibilities can be considerably extended in the
non-ideal case when using several strongest spectral peaks instead of a single one. This
extension occurs at the expense of a modest lowering of the nonlinear coupling strength
leading to a modest increase in the threshold intensity.

According to Fig.~2, phase matching is possible only for $j \geq
342$ above room temperature. Closer examination shows that tuning by
heating of the WGR becomes possible only owing to the impact of the
off-centering if the large radius $R$ exceeds noticeably the optimum
value of $1.54$~mm. Note lastly that the points of degeneracy, where
$\lambda_\mathrm{s}(T) = \lambda_\mathrm{i}(T)$, are almost
equidistant for the neighboring tuning curves with a temperature
step $\Delta T \approx 22^{\circ}$C.

In order to realize the extended OPO tunability, we have fabricated
a radially poled WGR with $2N = 686$, $R \simeq 1.55$~mm, and $r
\simeq 0.6$~mm. The corresponding $R/r$ ratio is known to be
favorable for coupling light in and out with a rutile (TiO$_2$)
prism~\cite{StrekalovOL09}. The resonator was fabricated from a
$500\,\mu$m thick $z$-cut wafer of congruent lithium niobate. The
domain structuring was performed using the standard
electric-field-poling technique with liquid
electrodes~\cite{Webjorn}; a radially patterned photo-resist
corresponded to the desired domain number. The quality of the
structuring was checked via domain selective etching -- more than
$50~\%$ of the desired domains are properly inverted. The radially
poled part of the wafer was cut out and shaped (diamond-turned and
polished) into a WGR with the desired $R/r$ ratio. The off-center
parameter $\delta R$ was estimated to be within $50\,\mu$m leading
to $\delta R/R \approx 0.03$.

Optical characterization of the fabricated WGR via line width and
free-spectral-range measurements has shown that the intrinsic
quality factor of the fundamental modes at $\lambda \simeq
1.04~\mu$m is $Q_\mathrm{p} \approx 4 \times 10^7$ and the large
radius $R \simeq 1.58$~mm. The last number is about $2.5~\%$ larger
compared to the desired one ($1.54$~mm); the difference is within
the accuracy of our fabrication procedure. Correspondingly, phase
matching is expected for $j \geq 350$ instead of $342$.

\begin{figure}[t]
\centering
\includegraphics[width=8cm]{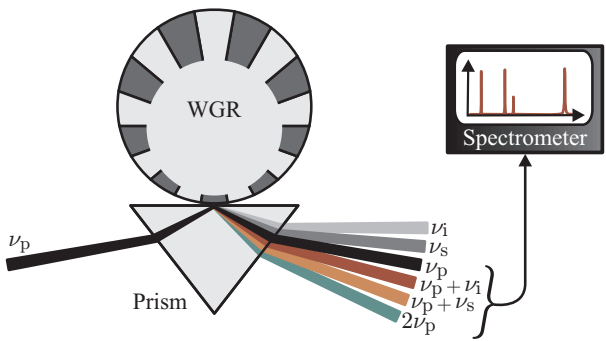}
\caption{Schematic of the WGR-OPO experimental setup. The frequencies of the interacting
waves are $\nu_\mathrm{p}$ (pump), $\nu_\mathrm{s}$ (signal), and $\nu_\mathrm{i}$
(idler).}\label{Fig3}
\end{figure}

The setup for our optical experiments, depicted in Fig.~3, comprises
the whispering gallery resonator mounted on a heatable post. The
latter one is used to change the resonator temperature up to
$62~^\circ$C. An extraordinarily polarized pump beam at $1.04~\mu$m
from an external-cavity diode laser is focused into a rutile prism.
In the focus, placed at the prism base, the pump beam couples into
the WGR via frustrated total internal reflection. In order to keep
the setup as simple and as stable as possible, even under
temperature expansion of the setup elements, the coupling prism
contacts the resonator rim. The residual pump wave and the generated
waves coupled out of the resonator are guided to a spectrometer
covering the wavelength range from $0.2$ to $1.1~\mu$m. The spectra
are collected for varying temperatures and pump powers.

As soon as the pump wave is coupled into the WGR, two narrow peaks
corresponding to the pump frequency $\nu_\mathrm{p}$ and to the
second harmonic $2\nu_\mathrm{p}$ become clearly visible in the
spectrum.
\begin{figure}[ht]
\centering
\includegraphics[width=8.1cm]{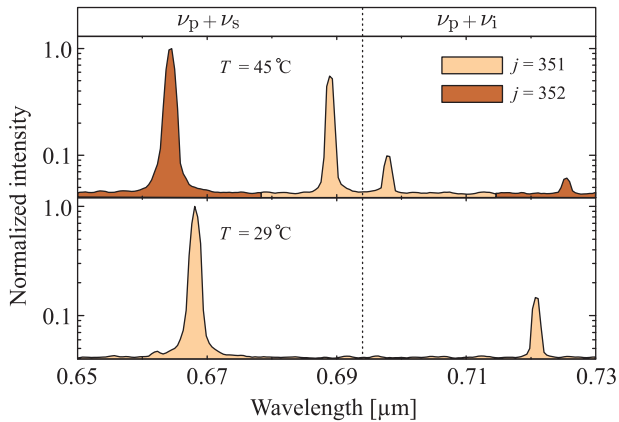}
\caption{Spectrum of the visible output of the WGR at a pump power of 10~mW for two
different temperature values showing sum frequency peaks at $\nu_\mathrm{p} +
\nu_\mathrm{s}$ and $\nu_\mathrm{p} + \nu_\mathrm{i}$. The corresponding parametric
processes are phase matched at $j = 351$ and $352$.}\label{Fig4}
\end{figure}
At pump powers $P$ exceeding a threshold value $P_{\rm th} \simeq 6$~mW, parametric
oscillation has been detected. In addition to the former spectral features, new peaks
arise well above the noise level, see Fig.~4. They can be reliably identified with the
sum frequencies $\nu_\mathrm{p} + \nu_\mathrm{s}$ and $\nu_\mathrm{p} + \nu_\mathrm{i}$,
where $\nu_\mathrm{s,i}$ are the signal and idler frequencies, such that $\nu_\mathrm{s}
+ \nu_\mathrm{i} = \nu_\mathrm{p}$. The accuracy of such measurements of
$\nu_\mathrm{i,s}$ is within the line width. In some cases, not only a single process,
but also two parametric processes, corresponding to adjacent values of $j$, are observed
simultaneously. While no direct measurements of the signal and idler waves were possible
with our spectrometer, the data obtained give a complete picture of the actual nonlinear
processes.

Changing the temperature $T$ between $20$ and $62\,^{\circ}$C, we
were able to tune the signal and idler wavelengths from $1.78$ to
$2.5~\mu$m, including the point of degeneracy $\lambda_\mathrm{s} =
\lambda_\mathrm{i}$, in steps of $(1-2)$~nm. The corresponding
experimental data are presented by the dots in Fig.~5.
\begin{figure}[ht]
\centering
\includegraphics[width=8.1cm]{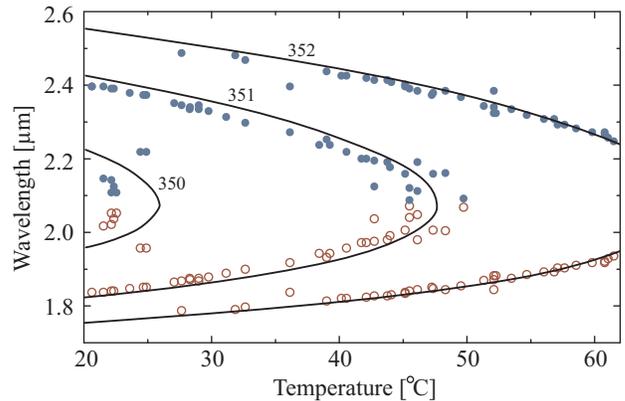}
\caption{Signal and idler wavelengths (open and filled dots)
evaluated from the frequency spectra as functions of the temperature
$T$. The solid lines are the tuning curves calculated numerically
for $j = 350$, $351$, and $352$ from Eq.~(2). }\label{Fig5}
\end{figure}
Most of the experimental dots nicely follow three tuning curves
(solid lines) calculated for the neighboring spectral peaks with $j
= 350$, $351$, and $352$. In accordance with theory, the measured
temperature difference between two adjacent points of degeneracy is
about $22~^{\circ}$C. Remarkably, filling of different tuning curves
with the dots is non-uniform. For $T \lesssim 30^{\circ}$C, the
curve with $j = 352$ is practically unfilled. The branches with $j >
352$ remain inactive in the whole temperature range.

Several issues, closely related to the above results, are worthy of discussion:

In some special cases, phase matching for $\chi^{(2)}$ nonlinear
processes can be realized in WGR's even without domain engineering,
i.e.\ in the single-crystal case. With lithium niobate, it is
possible for the processes involving modes of different
polarizations~\cite{FurstPRL10,Furst10}. However, the tunability is
restricted and the relevant nonlinear coefficient $d_{13}$ is about
five times smaller than the largest coefficient
$d_{33}$~\cite{Seres}.

Being focused on the tunability issue, we did not try to minimize
the pump threshold $P_{\rm th}$. While our experimental value
$P_{\rm th} \approx 6$~mW is already low, it can still be decreased
considerably. What are the prospects for decreasing the threshold
power? In order to estimate them, we write down the scaling relation
$P_{\rm th} \propto \nu_i\nu_s/Q^*_\mathrm{p} Q^*_\mathrm{s}
Q^*_\mathrm{i}\,d_j^2$, where $d_j = |S_j|d_{33}$ is the effective
nonlinear coefficient and $Q^*_\mathrm{p,s,i}$ are the loaded
quality factors incorporating the coupling losses~\cite{Matsko02}.
This relation contains the main parameters which can be varied in
the experiment. Because of the imperfect domain engineering, see
above, the actual reduction factors $|S_j| \approx 0.1$ were $(3-4)$
times smaller than they might be, which gives roughly one order of
magnitude increase in $P_{\rm th}$. About two orders of magnitude
increase stems from the overall coupling losses for
$\mathrm{p,s,i}$-waves -- bringing the coupling prism in contact
with the WGR rim noticeably decreases $Q^*_\mathrm{p,s,i}$. Thus,
the threshold power can be decreased by about three orders of
magnitude without sacrificing the tuning properties.

It is known that the phase-matching acceptance bandwidth increases
nearby the point of degeneracy, $\lambda_\mathrm{s} =
\lambda_\mathrm{i}$, if the $\mathrm{s,i}$-modes are of the same
polarization~\cite{Byer75}. This is why the experimental dots in
Fig.~5 are widely spread around the tuning curve for $j = 351$ at $T
\approx 45~^{\circ}$C. If a narrow acceptance bandwidth is required,
one can employ an OPO scheme with different
$\mathrm{s,i}$-polarizations. The domain number $2N$ must be
different in this case. About $1480$ domains are needed, instead of
former $2N \approx 700$, to realize this scheme for WGR radii
similar to the ones in our experiment. The structure of the tuning
curves is expected to be different as well. The above mentioned
indicates a high degree of flexibility of the nonlinear schemes
based on structured WGR's. By changing the domain number one can
proceed from broad-band to narrow-band gain and from identically
polarized to cross-polarized generated waves.

In conclusion, we have shown that radial poling of
whis\-per\-ing-gallery resonators made of lithium niobate allows to
combine a high tunability of nonlinear-optical processes, such as
optical parametric oscillation, with common low-power
continuous-wave light sources. The tuning characteristics differ
drastically from those known for bulk nonlinear schemes because of
the discrete geometry-dependent mode structure. The potential of
quasi-phase matching for shaping nonlinear processes is thus
strongly extended.

\vspace*{2mm}

We thank the DFG and the Deutsche Telekom AG for financial support.

\end{document}